\documentclass[twocolumn,showpacs,preprintnumbers,amsmath,amssymb,aps]{revtex4}

\usepackage{graphicx}
\usepackage{amssymb}
\usepackage{epstopdf}
\usepackage{bm}

\begin{document}
    
\title{General Relativity As an \AE ther Theory}
    
\author{Maurice J. Dupr\'e and Frank J. Tipler}
\affiliation{Department of Mathematics, Tulane University, New Orleans, LA 70118}
\date{\today}

\begin{abstract}
Most early twentieth century relativists --- Lorentz, Einstein, Eddington, for examples --- claimed that general relativity was merely a theory of the \ae ther.  We shall confirm this claim by deriving the Einstein equations using \ae ther theory.  We shall use a combination of Lorentz's and Kelvin's conception of the \ae ther.  Our derivation of the Einstein equations will not use the vanishing of the covariant divergence of the stress-energy tensor, but instead equate the Ricci tensor to the sum of the usual stress-energy tensor and a stress-energy tensor for the \ae ther, a tensor based on Kelvin's \ae ther theory.  A crucial first step is generalizing the Cartan formalism of Newtonian gravity to allow spatial curvature, as conjectured by Gauss and Riemann.
\end{abstract}

\maketitle

\section{Introduction}
Richard Feynman often emphasized the importance of having many
mathematically equivalent ways of expressing the same physical
theory.  In the lecture which he gave on the occasion of receiving
the 1965 Nobel Prize for physics, Feynman said: ``Theories of the
known, which are described by different physical ideas may be
equivalent in all their predictions and are hence scientifically
indistinguishable. However, they are not psychologically identical
when trying to move from that base into the unknown. For different
views suggest different kinds of modifications which might be made
and hence are not equivalent in the hypotheses one generates from
them in one's attempt to understand what is not yet understood. I,
therefore, think that a good theoretical physicist today might
find it useful to have a wide range of physical viewpoints and
mathematical expressions of the same theory (for example, of
quantum electrodynamics) available to him. This may be asking too
much of one man. Then new students should as a class have this.
$\ldots$ if my own experience is any guide, $\ldots$ if the
peculiar viewpoint taken is truly experimentally equivalent to the
usual in the realm of the known, there is always a range of
applications and problems in this realm for which the special
viewpoint gives one a special power and clarity of thought, which
is valuable in itself'' \cite{Feynman65}.

We shall follow Feynman and give a derivation of the Einstein field equations from \ae ther theory.  Most of the leading relativists in the early twentieth century, for examples Eddington \cite{Eddington1921} and even Einstein himself \cite{Einstein1920}, claimed that general relativity was an \ae ther theory, but they gave no mathematical demonstration of their claim. 

We shall provide the demonstration in this paper.  A huge number of \ae ther theories were proposed over the nineteenth century, and one could write a book describing them.  In fact, Edmund Whittaker wrote a {\it two} volume book (\cite{Whittaker1951}, \cite{Whittaker1954}) describing them.  All we shall need is two of these \ae ther theories, namely the theory of Lorentz, and the theory of Kelvin.

The first step is to generalize the Cartan theory of Newtonian gravity to allow spatial curvature.  To this curved space Cartan theory, we add Lorentzian \ae ther, which says the Maxwell equations {\it are} the theory of the \ae ther (\cite{Einstein1923}, p. 13).  We show that this implies that the curved space Poisson equation, $R_{tt} = 4\pi G\rho$ must become $R_{\mu\nu} =4\pi GS_{\mu\nu}$, where all components of the Ricci tensor $R_{\mu\nu} $ must be present.  

According to Einstein, in his {\it Autobiography} \cite{Einstein1959}, the most natural choice for the tensor $S_{\mu\nu}$ is the stress-energy tensor.  Einstein was uncomfortable with adding the term $-\frac{1}{2}g_{\mu\nu}R$ to the Ricci tensor, saying it was only introduced for``technical reasons,'' required by the vanishing of the covariant divergence of the stress-energy tensor.

Einstein was wise in being uncomfortable with this justification for adding this term. The modern view follows Noether and sees conservation laws as an expression of symmetries.  Energy conservation is a consequence of a timelike Killing vector, momentum conservation a consequence of an appropriate spacelike Killing vector, corresponding to invariance under spatial translation, and so on.  So the total energy need not be conserved in a spacetime with no timelike Killing field.  But $T{}^{\mu\nu}_{;\nu} = 0$ follows from $T{}^{\mu\nu}_{,\nu} = 0$ only by assuming the comma goes to semicolon rule.  The Noether theorems point out just how powerful an assumption this is.  The vanishing of  the divergence of the stress energy tensor is derived in Minkowski space using all the symmetries of Minkowski space.  But leaving Minkowski space for a general spacetime means losing the symmetries that allowed the derivation of $T{}^{\mu\nu}_{,\nu} = 0$ to start with!  

We shall avoid using $T{}^{\mu\nu}_{;\nu} = 0$ by assuming instead that the tensor $S_{\mu\nu}$ is a sum of two stress-energy tensors, the usual stress-energy tensor, and the stress-energy tensor for the \ae ther.  We shall show that the Lorentz theory, when combined with the Kelvin theory of the \ae ther gives a form for $S_{\mu\nu}$ such that the resulting theory is the familiar Einstein equations of general relativity.

In Section 2, we shall generalized the Cartan-Newton gravity theory to curved space, by generalizing the Misner axioms for Cartan-Newton gravity theory.  In Section 3, we then apply Lorentz-Kelvin \ae ther theory to obtain the Einstein field equations.  Finally in Section 4, we shall point out that all the basic ideas to construct the Einstein-\ae ther  field equations existed in the nineteenth century.  Using the PPN formalism, we shall show that the experimental evidence already existed in the nineteenth century to confirm this theory.

\bigskip
\centerline{\bf II. Generalizing the Misner Axioms for}
\centerline{\bf Newtonian Gravity}
\bigskip

What we shall do in this section is generalize Cartan's formulation to allow space to be curved.  To accomplish this we shall proceed by starting with a set of rigorous axioms for Newtonian gravity as curvature.  There are two such systems, one developed by Trautman and the other presented later by Misner.  These two systems are mathematically equivalent, but we shall use the Misner axioms, because Misner's system is much easier to generalize to the non-flat spatial case.  

The eight Misner axioms are given in Box 12.4 of MTW \cite{MTW}:

{\bf Axiom 1}: There exists a function $t$ called ``universal time,'' and a symmetric (i.e., torsion free) covariant derivative ${\bf \nabla}$ (with associated geodesics, parallel transport, curvature operator, etc.).

{\bf Axiom 2}: The 1-form ${\bf d}t$ is covariantly constant: 

\begin{equation}
{\bf \nabla_ud}t = 0
\label{eq:timeoneform}
\end{equation}
\noindent
for all vectors ${\bf u}$.

{\bf Axiom 3}:  Spatial vectors are unchanged by parallel transport around infinitesimal, closed curves, i.e., 

\begin{equation}
{\cal R}({\bf u},{\bf n}){\bf w} = 0
\label{eq:curvaturezero1}
\end{equation}
\noindent
if ${\bf w}$ is spatial for all vectors ${\bf u}$ and ${\bf n}$, and ${\cal R}{\bf(u,n)}$ is the curvature operator.

{\bf Axiom 4}:  All vectors are unchanged by parallel transport around infinitesimal, spatial, closed curves, i.e., 

\begin{equation}
{\cal R}({\bf v},{\bf w}) = 0
\label{eq:curvaturezero2}
\end{equation}
\noindent
for every spatial ${\bf v}$ and ${\bf w}$.

{\bf Axiom 5}:  The Ricci curvature tensor $R_{\alpha\beta} \equiv R^\mu_{\alpha\mu\beta}$ has the form

\begin{equation}
{\bf Ricci} = 4\pi\rho {\bf d}t\otimes {\bf d}t
\label{eq:RicciPoisson2}
\end{equation}
\noindent
where $\rho$ is the density of mass.

{\bf Axiom 6}: There exists a metric ``$\cdot$'' defined on spatial vectors only, which is compatible with the covariant derivative in the following sense:

\begin{equation}
{\bf \nabla_u(w\cdot v)} = {\bf( \nabla_uw)\cdot v} + {\bf( \nabla_uv)\cdot w}
\label{eq:metric2}
\end{equation}
\noindent
for any spatial vectors ${\bf w}$ and ${\bf v}$, and for any ${\bf u}$ whatsoever.

{\bf Axiom 7}:  The Jacobi curvature operator ${\cal J}{\bf (u.n)}$, defined for any vectors ${\bf u}$, ${\bf n}$, and ${\bf p}$ by 

\begin{equation}
{\cal J}{\bf (u,n)p} = \frac{1}{2}\left[{\cal R}{\bf(p,n)u} + {\cal R}{\bf(p,u)n}\right] 
\label{eq:Jacobicurvature}
\end{equation}
\noindent
is ``self-adjoint'' when operating on spatial vectors, i.e.,

\begin{equation}
{\bf v \cdot}\left[{\cal J}{\bf (u,n)w}\right] = {\bf w \cdot}\left[{\cal J}{\bf (u,n)v}\right]
\label{eq:Jacobicurvatureselfadjoint}
\end{equation}
\noindent
for all spatial vectors ${\bf v}$ and ${\bf w}$, and for any vectors ${\bf u}$ and ${\bf n}$ whatsoever.

{\bf Axiom 8}: Ideal rods measure the lengths that are associated with the spatial metric, and ideal clocks measure universal time $t$ or some multiple thereof.  Furthermore, freely falling particles move along geodesics of ${\bf \nabla}$.

Let us remind the reader what these axioms are intended to accomplish on the connection: (1) ensure that the only non-vanishing components are $\Gamma^i_{tt}$; (2) ensure that the spatial vector $\Gamma^i_{tt}$ is the gradient of some scalar field, i.e., $\Gamma^i_{tt} = \phi_{,i}$, (3) $\Gamma^i_{kl} = 0$, so that (4) the spatial metric is the metric of flat space.

We want to generalize these axioms so that the following is true on the connection: (1) ensure that the only non-vanishing components of the connection are $\Gamma^i_{tt}$ and $\Gamma^i_{kl}$; (2) ensure that the spatial vector $\Gamma^i_{tt}$ is still the gradient of some scalar field, and (3) ensure that $\Gamma^i_{kl}$ arises from a spatial Reimannian metric.  For simplicity, we shall assume that in what follows, all vectors are written locally in a coordinate frame basis.

We can accomplish this by deleting Axiom 4 (which imposes spatial flatness), and replacing Axiom 3 by

{\bf Axiom 3A} The basis vector ${\bf e_t}$ dual to the 1-form ${\bf d}t$ (that is, $<{\bf d}t, {\bf e_t}> = 1$), is itself covariantly constant:

\begin{equation}
{\bf \nabla_we_t} = 0
\label{eq:timeoneformdual}
\end{equation}
\noindent
at least for all spatial vectors ${\bf w}$.

Axiom 1 implies that $\Gamma^t_{\alpha\beta} = 0$ for all $\alpha$ and $\beta$.  It is easily checked that Axiom 3A implies that $\Gamma^i_{t j} = \Gamma^i_{j t} = 0$, leaving us with $\Gamma^i_{tt}$ and $\Gamma^i_{kl}$ as the only non-vanishing connection coefficients.  In such a case, Misner's Axiom 6 will force the spatial components of the connection, $\Gamma^i_{kl}$ to arise from a spatial Riemannian metric. 

Axiom 2 and $<{\bf d}t, {\bf e_t}> = 1$ implies that ${\cal R}{\bf(w,e_t)e_t}$ is spatial for ${\bf w}$ spatial.  Then Axiom   7 implies that $R_{itkt} = R_{ktit}$.  Writing $\Gamma^i_{tt} \equiv v^i$), this gives:

\begin{equation}
v_{i;k} = R_{itkt} = R_{ktit} =  v_{k;i}
\label{eq:Ritkt}
\end{equation}
\noindent
But this implies

\begin{equation}
v_{i;k} - v_{k;i} = v_{i,k} - v_{k,i} = 0
\label{eq:curlvanishes}
\end{equation}
\noindent
since the covariant curl equals the curl.  Thus the vector field $v^i = \Gamma^i_{tt} = g^{ij}v_j$ is the gradient of a scalar field, since its curl vanishes.  We also have

 \begin{equation}
R_{tt} = R^i_{tit} = \Gamma^i_{tt;i} = g^{ij}\left[\frac{\partial^2\phi}{\partial x^i\partial x^j} - \Gamma^k_{ij}\frac{\partial\phi}{\partial x^k}\right] = \nabla^2\phi
\label{eq:LaplaceBelatrami}
\end{equation}
\noindent
which is just the Laplace operator for curved space acting on the scalar field $\phi$.  

Thus, Axiom 5 gives the required generalization of Poisson's equation to curved space.

Of course, there is a simpler approach if one does not insist on expressing Newtonian gravity in terms of Poisson's equation, but  instead in terms of the spatial vector $v^i$.  Then one does not need Axiom 3A at all, and one has, instead of (\ref{eq:LaplaceBelatrami}), simply

 \begin{equation}
R_{tt} = R^i_{tit} = \Gamma^i_{tt;i} =  {\vec\nabla}\cdot{\vec v}
\label{eq:LaplaceBelatrami2}
\end{equation}

This approach, where one does not require that $v^i = \nabla^i\phi$, is sufficient to develop the Milne-McCrea Newtonian cosmology \cite{Tipler96}.

It is important to note that the curved space version of Cartan-Newton theory has to take the spatial metric as a given. There are no equations to determine the spatial metric.  So mathematically, one is allowed to impose the spatial metric arbitrarily, and then use the matter distribution and boundary conditions to determine the potential, or the vector field $v^i$.  Einstein's theory is thus more restrictive, because the metric is constrained by the ten Einstein equations, rather than the single curved space Poisson equation.

This arbitrary background geometry is a ``prior geometry'' in the words of MTW (section 17.6).  As MTW emphasize, the requirement that there is no ``prior geometry''--- that the metric is entirely determined by the field equations for gravity --- actually fathered general relativity.  There being no freedom in geometry is just one manifestation that general relativity is more restrictive than Newtonian gravity.   A second manifestation is that in Newtonian gravity, the connection is not metric, but a more general affine connection.  A third manifestation we shall see in the next section, that general relativity follows only from a very special \ae ther theory.

Georg F. B. Riemann gave a famous lecture \cite{Clifford}, ``On the Hypotheses Which Lie at the Bases of Geometry,'' where he proposed that the
universe might be spatially a three-sphere rather than Euclidean
three-space.  It is well-known that Karl F. Gauss was in
the audience, and was enthusiastic about  the lecture --- in fact,
Gauss himself chose the topic of Riemann's lecture.  A
translation of Riemann's lecture was made for {\it Nature} by William K. Clifford, who wrote in full approval of Riemann's lecture.  So Riemann, Gauss, and Clifford  --- three of the greatest
mathematicians of the nineteenth century ---  believed that a
three-sphere was the preferred topology for the universe, as Einstein believed later, and Dante much earlier \cite{Peterson1979}..

There is, however, a serious difficulty with a three-sphere universe in Newtonian gravity theory.  If mass is positive, then Poisson's equation does not allow  a three-sphere
universe.  To see this,
integrate $\nabla^2\Phi= 4\pi G\rho$ over the entire three-sphere,
getting  $(1/(4\pi G))\int \rho{\sqrt g}d^3x =  \int \nabla^2\Phi
{\sqrt g}d^3x = \int div\, grad\Phi {\sqrt g}d^3x = \int_S
grad\Phi \,dS = 0$, the last two steps using the Gauss Divergence
theorem and the fact that the last integral is over the boundary
$S$ of the three-sphere, and thus this integral is zero since the
three-sphere has no boundary.  But $\int \rho{\sqrt g}d^3x = 0$ is
impossible if $\rho \geq 0$, unless $\rho = 0$ everywhere.  This fact may be the reason why Gauss and Riemann did not develop their idea that physical space was curved. 

Had they done so, general relativity would have had an easier time being accepted.     Recall that one of the main objections to general relativity was based on Occam's Razor, namely that general relativity depended on ten potentials $g_{\mu\nu}$ rather than the single Newtonian potential $\Phi$.  Had the Poisson equation for curved space been considered the Newtonian gravity equation, then physicists would have realized that Newtonian gravity theory required the determination of seven potentials: $\Phi$ and the six components of the spatial metric $g_{ij}$ with the latter six being undetermined by the boundary conditions.  Ten potentials is not significantly greater than seven, and further, the Einstein theory provides an additional nine equations, allowing the boundary conditions to yield a unique solution.

\bigskip
\centerline{\bf III. Proof that the Einstein Gravity Equations}
\centerline{\bf are a Special Case of the Newtonian Gravity}
\centerline{\bf Equations Coupled to a Luminiferous \AE ther}
\bigskip

A central point of Lorentz's 1904 paper, in which he derived the Lorentz transformations, was that the Maxwell equations --- for Lorentz, the equations of the \ae ther --- do not allow an absolute time to be defined.  This is of course now obvious since the speed of light in the vacuum is a constant, independent of a inertial observer.  So the \ae ther can be thought of as defining a time direction different from what we may have thought of as Newtonian absolute time.  Trautman showed \cite{Trautman1966} that this time direction can be defined as a 4-dimensional vector $u^\mu $, which he called a ``rigging,'' with

\begin{equation}
u^\mu \equiv \left(u^t,\, u^x,\, u^y,\, u^z \right)
\label{eq:uvector}
 \end{equation}
\noindent where the spatial components are non-zero, but
constant over all space for an inertial observer.  For a general observer, all components are functions of space and time.  We shall impose  the constraint that $(u^t)^2 > (u^x)^2 + (u^y)^2 +
(u^z)^2$, so that it will be ``timelike," in the sense that the
component in the time direction is larger than any space
direction.  The components $u^j$ can be viewed as the
components of the velocity of the \ae ther with respect to Newtonian time in the $j^{th}$
direction, and we will also set $u^t = c$.

The vector field $u^\mu$ defines a 4-dimensional metric:

\begin{equation}
g^{\mu\nu} \equiv g^{ij} -\frac{u^\mu u^\nu}{c^2}
\label{eq:Lorentzmetric}
 \end{equation}

\noindent
where $g^{ij}$ is the 3-dimensional  spatial metric in Section II, written as a 4-metric in which all non-spatial components are zero.  The symmetric rank two tensor  $g_{\mu\nu}$ defines a 4-D Lorentz metric.

If space is not spatially flat, then the spatial Riemannian metric will define a metric connection, and we {\it might} thus have {\it two} connections, one from the spatial metric, and one in the time direction only.  

But having the 4-metric $g_{\mu\nu}$ means  that there is no longer a ``natural' division between time and space, and hence there is no natural division between the purely timelike connection and the purely spacelike connection.  Since the rigging defines a pseudo-Riemannian metric, it is natural --- but not required ---to assume that the entire connection arises from the pseudo-Riemannian metric.  We emphasize that this is an added constraint on the full \ae ther theory, which would in principle have two connections, one from Newtonian gravity, and yet another from the pseudo-Riemannian geometry.  We suspect, but do not attempt to prove, that maintaining the distinction between two such connections would be very difficult.  

Essentially, the requirement that the connection arise entirely from the metric is nothing but the ``no prior geometry'' assumption, which, as we pointed out earlier, is the only assumption that will allow the geometry to be determined by the matter distribution and the boundary conditions.  Once again, MTW have emphasized that the  ``no prior geometry'' assumption is the basic assumption of general relativity.  It is also an essential assumption of the curved \ae therial Newtonian gravity theory we develop here.  In effect, we use it to require that the only connection is the metric connection, and assuming, like Cartan-Newton and Einstein, that particles move along geodesics.  The geodesics are necessarily those of the metric connection, since there is no other connection

Since the existence of the \ae ther by definition tells us that
Newtonian time cannot be unique, the time index $t$ in the
Cartan-Poisson equation must be replaced with a pair of time
indices $t$ and $t'$:

\begin{equation}
R_{tt'} =  4\pi G\rho'
\label{eq:minimalNewton}
 \end{equation}

\noindent where $\rho'$ is some density appropriate to this pair
of indices.  But (\ref{eq:minimalNewton}) is really a
tensor equation of the form

\begin{equation}
R_{\mu\nu} =  4\pi GS_{\mu\nu}
\label{eq:generalEinstein}
 \end{equation}

\noindent because the LHS of (\ref{eq:minimalNewton}) are
components of a tensor, and further, if two symmetric tensors
agree for all possible $t'$ time coordinates, they are the same
tensor in all space and time dimensions.  This last statement is Proposition 3.3.4 of Sachs and Wu ((\cite{SachsWu} , p. 72).

The question is, what should we select for the tensor $S_{\mu\nu}$.  According to Einstein in his {\it Autobiography}: ``On the right side [of the Einstein equations] we shall  then have to place a tensor also in place of [the mass density] $\rho$.  Since we know from the special theory of relativity that the (inertial) mass equals energy, we shall have to put on the right side the tensor of energy-density --- more precisely the entire energy-density, insofar as it does not belong to the pure gravitational field (\cite{Einstein1959}, p. 75.).

We propose to follow Einstein exactly: the tensor $S_{\mu\nu}$ must be the entire stress-energy tensor ``insofar as it does not belong to the pure gravitational field.'' Since by hypothesis, we have an \ae ther, we must include the \ae ther stress energy:

\begin{equation}
R_{\mu\nu} =  4\pi GS_{\mu\nu} =4\pi G\left(T_{\mu\nu} + T^{{\rm aether}}_{\mu\nu}\right)
\label{eq:generalEinstein2}
\end{equation}

\noindent 
where $T_{\mu\nu}$ is the tensor for the energy density
of ordinary ponderable matter, and $T^{{\rm aether}}_{\mu\nu}$ is
the energy density of the \ae ther.  We shall now show that the \ae ther theory of Lorentz and Kelvin gives the form of  $T^{{\rm aether}}_{\mu\nu}$.

We start with the result, originally derived by Maxwell in 1873
(\cite{Maxwell1873}, section 792; p. 391 of volume II) that for an electromagnetic wave
traveling in the $i$th direction, where $i$ is either $x$, $y$ or
$z$ in the \ae ther, then $\rho = p_i/c^2$ where $\rho$ is the
``mass'' density of the electromagnetic radiation, and the $p_i$
are the pressures in the $i$th direction.  (Lorentz had derived $E=mc^2$ for electromagnetic fields.)

This gives the general relation between the density and the pressure of an electromagnetic wave:

\begin{equation}
T_{tt} = \rho_{\rm EM} = (p^{\rm EM}_x + p^{\rm EM}_y + p^{\rm EM}_z)/c^2 = T^i_i/c^2
\label{eq:EMmasspressure}
\end{equation}

Notice that we are equating the matter density to the trace of the spatial part of the stress-energy tensor; we need not assume that $T^{\mu\nu}$ can be diagonalized, just that $T^\mu_\mu = 0$.  Since the Lorentz \ae ther theory requires the Maxwell equation to be the equations for the \ae ther,   this means that equation (\ref{eq:EMmasspressure}) must also be the equation for the corresponding quantities for the \ae ther; there is nothing else:

\begin{equation}
\rho_{\rm aether} = (p^{\rm aether}_x + p^{\rm aether}_y + p^{\rm aether}_z)/c^2
\label{eq:aethermasspressure}
\end{equation}

Now we use the Kelvin theory of the \ae ther, in which {\it all} pressures are ultimately due to the \ae ther.  Kelvin devoted an entire book \cite{Kelvin1904} to various models for how to accomplish this, but for our purposes the details don't matter. We also note that Lorentz had the same hope, to reduce all phenomena, in particular pressures of all types, to \ae ther phenomena.

So if all pressures are ultimately \ae ther pressures, the most natural way to express this is to simply delete the superscript ``\ae ther'' in the pressures in equation (\ref{eq:aethermasspressure}):

\begin{equation}
\rho_{\rm aether} = (p_x + p_y + p_z)^/c^2
\label{eq:centraleaetherquation}
\end{equation}

Equation (\ref{eq:centraleaetherquation}) makes rigorous Kelvin's belief \cite{Kelvin1901} that the \ae ther must generate gravity, but that it cannot do so in the absence of matter.

Equation (\ref{eq:centraleaetherquation}) implies $T^{{\rm aether}}_{\mu\nu} = T_{\mu\nu} -
g_{\mu\nu}g^{\alpha\beta}T_{\alpha\beta}$.  To see this, choose coordinates locally so that $g_{\alpha\beta} =
\eta_{\alpha\beta} $, so that in particular $g_{tt} = -1$,  and
thus

\begin{eqnarray}
\rho^{aether} \equiv T^{aether}_{tt} = (p_x + p_y + p_z)/c^2\nonumber\\
 = T_{tt} + [-T_{tt}  + (p_x + p_y + p_z)/c^2]\nonumber\\
 = T_{tt} - g_{tt} [-T_{tt}  + (p_x + p_y + p_z)/c^2]\nonumber\\
 = T_{tt} - g_{tt}  g^{\alpha\beta}T_{\alpha\beta} = T_{tt} - g_{tt} T
\label{eq:contractT}
\end{eqnarray}

\noindent where we have written the density of ordinary matter as
$T_{tt}$.  In other words, if there are no labels to the tensor
$T$ it is the tensor with only non-\ae ther material.  This yields

\begin{equation}
T^{\rm aether}_{\mu\nu}t^\mu t^\nu =  T_{tt} - g_{tt} T = (T_{\mu\nu} - g_{\mu\nu} T)t^\mu t^\nu
\label{eq:contractT2}
\end{equation}

Hence, we have an equality between two tensors for all timelike
unit vectors.  But recall that this equality implies the equality of the tensors
themselves:

\begin{equation}
 T^{\rm aether}_{\mu\nu} = T_{\mu\nu}  -T g_{\mu\nu}
\label{eq:aethertensor}
\end{equation}

This completes the derivation of the energy tensor for the \ae
ther, and thus derives the Einstein field equations as the
Newtonian equations for gravity in which the gravitating \ae ther
is included.

Notice that the \ae ther explains, without  using the weak field limit, why the the constant $4\pi$ in Poisson's equation is replaced by $8\pi$ in the Einstein equations, .  The \ae ther also enforces $T^\mu_{;\mu} =0$ without using the comma goes to semicolon rule.  Thus the \ae ther also implies $T^\mu_{,\mu} =0$ in Minkowski space, a law that would have been grossly violated \cite{LindblomNester1975} if we had set $S_{\mu\nu} = T_{\mu\nu}$. 

\bigskip
\centerline{\bf IV. Conclusion}
\bigskip

The PPN formalism (MTW, chapter 39) can be used to see the relative effects of time curvature, space curvature, and the fact that the Maxwell \ae ther combines these into a four-metric theory of gravity.  Recall that in isotropic coordinates, the PPN metric for a spherically symmetric field is  

\begin{equation}
ds^2 = g_{tt}c^2dt^2 + g_{ss}[dr^2 + r^2(d\theta^2 + \sin^2\theta d\phi^2)]
\label{eq:PPNmetric}
\end{equation}

\noindent
where
\begin{equation}
g_{tt} =  -\left[1-2\frac{GM}{rc^2} + 2\beta\left(\frac{GM}{rc^2}\right)^2\right]
\label{eq:gtt}
\end{equation}

\noindent
and
\begin{equation}
g_{ss} =  \left[ 1 + 2\gamma\frac{GM}{rc^2}\right]
\label{eq:gss}
\end{equation}

For general relativity, the PPN parameters have values $\beta = \gamma =1$.  For a solar system experiment, $M$ is the mass of the Sun $M_\odot$.  The angle $\Delta\alpha$ of deflection by a light ray passing by the Sun is

\begin{equation}
\Delta\alpha = \frac{1}{2}(1 + \gamma)\frac{(GM_\odot/c^2)}{b}
\label{eq:lightdeflection}
\end{equation}

\noindent
where $b$ is the light ray's impact parameter, and $GM_\odot/b = 1.75"$ if the light ray just grazes the limb of the Sun.

For a planet in an elliptical orbit around the Sun, with semimajor axis $a$ and eccentricity $e$, the perihelion shifts forward by an angle $\Delta\phi_0$ with each circuit around the ellipse given by

\begin{equation}
\Delta\phi_0 = \left[\frac{2-\beta +2\gamma}{3}\right]\left[\frac{6\pi}{1-e^2}\right]\left[\frac{GM_\odot}{ac^2}\right]
\label{eq:perihelionshift}
\end{equation}

\noindent
which for Mercury is $42.98 \pm 0.04$ seconds of arc per century \cite{Will2006}.

The parameter $\gamma$ measures the deviation of the spatial metric from flat space, and (\ref{eq:lightdeflection}) shows that fully half the value of the deflection of a light ray comes from the spatial curvature; the other half comes from the purely Newtonian curvature in the time direction.  So without allowing for spatial curvature, one half of the actual light deflection would be unaccounted for.

In our curved Newtonian spacetime, without the \ae ther, one can obtain the observed Einstein shift by putting in the necessary spatial curvature by hand, basically as a fudge factor since the spatial metric in the curved space Poisson equation is entirely arbitrary.  The ``no prior geometry'' assumption makes the spatial curvature non-arbitrary, and in fact gives exactly the observed value.  So from the \ae ther theory point of view, the light deflection experiment is testing the ``no prior geometry'' assumption.

In spatially flat, non-\ae ther Newtonian theory, $\beta = \gamma =0$, so the PPN formula, valid for all metric theories of gravity, shows us that there would be a perihelion shift of Mercury 2/3 that of general relativity in any metric theory.  That is, the number 2 in the numerator of the first factor in brackets in (\ref{eq:perihelionshift}) is due to a special relativistic effect that would be present even if $\beta = \gamma = 0$.

To see this, recall that special relativistic effects are of order $v^2/c^2$ in the limit $v\ll c$.  Now the perihelion shift is a deviation from an assumed Kepler orbit, for which $GM_\odot/a = v^2$ when $e \ll 1$. So the last factor in brackets in (\ref{eq:perihelionshift}) is just $v^2/c^2$, exactly what we would expect for a special relativistic effect.  To take the full spatially flat, non-\ae ther limit for the perihelion shift, one must not only set $\beta = \gamma =0$, but also take the limit $v\rightarrow 0$, which means taking the limit $a\rightarrow \infty$ in (\ref{eq:perihelionshift}).  Thus, just having an \ae ther, which requires a rigging, which in essence is just special relativity, would give 2/3 of the perihelion shift.  

Lord Kelvin \cite{Kelvin1911} wrote an article in 1859 on LeVerrier's discovery of Mercury's perihelion shift, agreeing incorrecty with LeVerrier that it was probably due to a series of small subMercurian planets.  Remarkably, it was actually due to the gravitational effect of Kelvin's \ae ther! 

Equally remarkable is the fact that LeVerrier's 1859 number for the periherion shift, $38''$ per century, which is within 12\% of the actual value of $43''$ per century, is sufficiently accurate to see not only the spatial curvature correction factor $\gamma$, but also the $\beta$ factor correction to the curvature in the time direction.  The $\beta$ factor is to be regarded as an ``\ae ther'' correction to the gravitational field, since it is a consequence of $R_{\mu\nu} = 0$ rather than the non-\ae ther $R_{tt} = 0$.  Setting $\beta = \gamma =0$ gives $(2/3)43'' =29''$ which is 24\% lower than LeVerrier's $38''$.  Setting only $\beta = 0$, but keeping the spatial curvature at its full general relativistic $\gamma = 1$, gives $(4/3)43'' =54''$ which is 42\% higher than LeVerrier's $38''$.  Since Leverrier was only 12\% off the true value, he would have noticed either deviation from the true value.  So the nineteenth century physicists were not only conceptually capable of deriving general relativity, but they had by 1859 the data that would confirm the \ae ther theory that is general relativity.

We began with Feynman, let us end with Feynman.  In his Nobel Prize lecture, Feynman said he wondered what Dirac meant by saying two mathematical expressions were ``analogous'' to each other.  Feynman calculated that ``analogous''  meant ``equal.''
In his {\it Autobiography},  Einstein wrote: ``In the case of the relativistic theory of the gravitational field,  $R_{\mu\nu}$ takes the place of $\nabla^2\Phi$ (\cite{Einstein1959}, p. 73).  Cartan showed that for Newtonian theory in flat space, $R_{tt}$ was actually equal to $\nabla^2\Phi$.  In this paper, we have extended this equality to the case of curved space.  Einstein went on to make the remark we quoted earlier, ``On the right side [of the Einstein equations] we shall  then have to place a tensor also in place of [the mass density] $\rho$.  Since we know from the special theory of relativity that the (inertial) mass equals energy, we shall have to put on the right side the tensor of energy-density --- more precisely the entire energy-density, insofar as it does not belong to the pure gravitational field.''

In this paper we have shown that Lorentz \ae ther theory requires that all times $t$ must be permitted for $R_{tt}$, and that this implies that the entire tensor $R_{\mu\nu}$ must describe gravity.  We showed that the spatial metric and the Trautman vector field $u^\mu$ give a 4-metric, and that ``no prior geometry implies that the entire connection must be the 4-metric connection. Finally, we showed that, following Einstein, if the right side is the {\it entire} energy density --- the usual entire energy density plus the entire energy density of the Lorentz-Kelvin \ae ther --- then the full Einstein equations are obtained.

\par
{\bf Acknowledgments}:  We are grateful to Prof. C.M. Will for a discussion on the correct Newtonian limit in the PPN formalism.

\end{document}